\documentclass[12pt]{article}

\usepackage[dvipdfmx]{graphicx}
\usepackage{amsfonts,amsmath,amssymb}
\usepackage[mathscr]{eucal}
\usepackage{ascmac}
\usepackage{dcolumn}
\usepackage{bm}
\usepackage[colorlinks=true,linkcolor=blue,citecolor=blue]{hyperref}
\usepackage{hyperref}
\usepackage{color}

\begin{document}


\newcommand{\vev}[1]{ \left\langle {#1} \right\rangle }
\newcommand{\bra}[1]{ \langle {#1} | }
\newcommand{\ket}[1]{ | {#1} \rangle }
\newcommand{\EV}{ \ {\rm eV} }
\newcommand{\KEV}{ \ {\rm keV} }
\newcommand{\MEV}{\  {\rm MeV} }
\newcommand{\GEV}{\  {\rm GeV} }
\newcommand{\TEV}{\  {\rm TeV} }
\newcommand{\1}{\mbox{1}\hspace{-0.25em}\mbox{l}}
\newcommand{\Red}[1]{{\color{red} {#1}}}

\newcommand{\lmk}{\left(}  
\newcommand{\rmk}{\right)}
\newcommand{\lkk}{\left[}  
\newcommand{\rkk}{\right]}
\newcommand{\lhk}{\left \{ }  
\newcommand{\rhk}{\right \} }
\newcommand{\del}{\partial}  
\newcommand{\la}{\left\langle} 
\newcommand{\ra}{\right\rangle}
\newcommand{\half}{\frac{1}{2}}

\newcommand{\bea}{\begin{array}}
\newcommand{\eea}{\end{array}}
\newcommand{\beq}{\begin{eqnarray}}
\newcommand{\eeq}{\end{eqnarray}}

\newcommand{\dd}{\mathrm{d}}
\newcommand{\Mpl}{M_{\rm Pl}}
\newcommand{\mg}{m_{3/2}}
\newcommand{\abs}[1]{\left\vert {#1} \right\vert}
\newcommand{\mphi}{m_{\phi}}
\newcommand{\Hz}{\ {\rm Hz}}
\newcommand{\for}{\quad \text{for }}
\newcommand{\Min}{\text{Min}}
\newcommand{\Max}{\text{Max}}
\newcommand{\Kahler}{K\"{a}hler }
\newcommand{\cphi}{\varphi}
\newcommand{\Tr}{\text{Tr}}
\newcommand{\diag}{\text{diag}}

\newcommand{\SUf}{SU(N_f)_{\rm f}}
\newcommand{\Upq}{U(1)_{\rm PQ}}
\newcommand{\Zpq}{Z^{\rm PQ}_3}
\newcommand{\Cpq}{C_{\rm PQ}}
\newcommand{\Ndw}{N_{\rm DW}}
\newcommand{\Fpq}{F_{\rm PQ}}
\newcommand{\fpq}{v_{\rm PQ}}
\newcommand{\Br}{{\rm Br}}
\newcommand{\Lag}{\mathcal{L}}
\newcommand{\Lqcd}{\Lambda_{\rm QCD}}
\newcommand{\LQCD}{\Lambda_{\rm QCD}}

\def\lrf#1#2{ \left(\frac{#1}{#2}\right)}
\def\lrfp#1#2#3{ \left(\frac{#1}{#2} \right)^{#3}}


\begin{titlepage}

\baselineskip 8mm

\begin{flushright}
TU-1001\\
IPMU 15-0117
\end{flushright}

\begin{center}

\vskip 1.2cm

{\Large\bf
Strongly~broken~Peccei-Quinn~symmetry in~the~early~Universe 
}

\vskip 1.8cm

{\large Fuminobu Takahashi$^{a,b}$ 
and Masaki Yamada$^{b,c}$ 
}

\vskip 0.4cm

{\it$^a$Department of Physics, Tohoku University, 
Sendai, Miyagi 980-8578, Japan}\\
{\it$^b$Kavli IPMU (WPI), TODIAS, The University of Tokyo, 
Kashiwa, Chiba 277-8583, Japan}\\
{\it$^c$
Institute for Cosmic Ray Research, 
ICRR, The University of Tokyo,\\
Kashiwa, Chiba 277-8582, Japan}

\date{\today}
\vspace{2cm}

\begin{abstract}  
We consider QCD axion models where the Peccei-Quinn symmetry is badly broken
by a larger amount in the past than in the present,  in order to avoid the axion isocurvature problem. 
Specifically we study supersymmetric axion models where the Peccei-Quinn symmetry is dynamically 
broken by either hidden gauge interactions or the $SU(3)_c$ strong interactions whose
 dynamical scales are temporarily enhanced by the dynamics of flat directions.
The former scenario predicts  a large amount of self-interacting dark radiation as the hidden gauge symmetry is weakly
coupled in the present Universe. We also show that the observed amount of baryon asymmetry can be generated
by the QCD axion dynamics via  spontaneous baryogenesis. We briefly comment on the case in which
the PQ symmetry is broken by a non-minimal coupling to gravity.

\end{abstract}

\end{center}
\end{titlepage}

\baselineskip 6mm


\section{Introduction
\label{sec:introduction}}
The strong CP phase is constrained to be less than $10^{-10}$ by the neutron EDM experiments~\cite{Baker:2006ts}, and
the problem of why it is so small is known as the strong CP problem. 
One of the solutions is the Peccei-Quinn (PQ) mechanism~\cite{Peccei:1977hh, Peccei:1977ur}, 
which renders the strong CP phase a dynamical variable, an axion, which
arises as a pseudo Nambu-Goldstone (NG) boson 
in association with the spontaneous breakdown of the global PQ symmetry~\cite{Weinberg:1977ma}. 
The axion acquires a mass through nonperturvative effects of QCD~\cite{'tHooft:1976up, 'tHooft:1976fv} 
and its vacuum expectation value (VEV) cancels the undesired CP phase. The dynamical relaxation necessarily induces
coherent oscillations of the axion, which can account for
dark matter in the Universe~\cite{Preskill:1982cy, Abbott:1982af, Dine:1982ah}.

The PQ symmetry must be of extremely high quality to solve the strong CP problem, as even a tiny breaking 
would results in a too large CP phase in contradiction with the 
experiments~\cite{Georgi:1981pu,Lazarides:1985bj,Dine:1986bg,Kamionkowski:1992mf, Holman:1992us,Barr:1992qq,Dobrescu:1996jp}. 
The high quality of the PQ symmetry could be due to a non-trivial discrete symmetry of high order $Z_N$~\cite{Dine:1992ya} or
a higher dimensional gauge symmetry (or its combination with an anomalous U(1)$_A$ symmetry)~\cite{Svrcek:2006yi}.%
\footnote{
See Refs.~\cite{Barr:2001vh, Izawa:2002qk, Izawa:2004bi, Harigaya:2015soa, Kawasaki:2015lea} for 
other models of the origin of the PQ symmetry. 
}
Here we would like to emphasize that, from a purely phenomenological point of view, the PQ symmetry could be 
badly broken by a larger amount in the early Universe, as long as the extra breaking disappears in the present Universe. 
Interestingly, in the presence of such a large PQ breaking in the early Universe,
the axion becomes so heavy that its quantum fluctuations are suppressed, avoiding tight constraints on the axion isocurvature 
perturbations~\cite{Dine:2004cq,Jeong:2013xta, Higaki:2014ooa, Dine:2014gba, Choi:2015zra}.\footnote{
The stronger QCD was proposed to suppress the axion abundance~\cite{Dvali:1995ce}.
See also Ref.~\cite{Choi:1996fs,Banks:1996ea}.
There are various ways to suppress the axion isocurvature perturbations~\cite{Linde:1990yj, Linde:1991km, Kasuya:1996ns, 
Kasuya:1997td, Folkerts:2013tua,  Kawasaki:2013iha, Nakayama:2015pba, Kawasaki:2015lea, Harigaya:2015hha}.}

In this paper we study a possibility that the PQ symmetry is largely broken temporarily in the early Universe, 
and discuss their cosmological implications based on a couple of models.
Specifically we first discuss supersymmetric axion models in which the PQ symmetry is largely broken 
by non-Abelian hidden gauge interactions which are strongly coupled during inflation.
If the axion acquires a sufficiently heavy mass,  its quantum fluctuations would be suppressed. 
 In the present Universe, on the other hand, the hidden gauge interactions must be weakly coupled,
which is made possible by a dynamics of flat directions in the hidden sector. 
We shall also see that the axion dynamics after inflation can generate a sizable baryon asymmetry {\it a la} spontaneous
baryogenesis~\cite{Cohen:1987vi,Dine:1990fj,Cohen:1991iu,Kusenko:2014uta,Daido:2015gqa}.
We also show that the suppression can be realized in a more economical model where the PQ symmetry is broken only by
the $SU(3)_c$ strong  interactions. The QCD scale can be enhanced if the $H_u H_d$ flat direction 
acquires a large VEV in the early Universe~\cite{Dvali:1995ce,Jeong:2013xta}.

The rest of the paper is organized as follows. 
In Sec.~\ref{sec:1},  we briefly review observational constraints on the axion isocurvature perturbations.
In Sec.~\ref{sec:model2}, 
we provide supersymmetric axion models where the dynamical scales of the hidden or $SU(3)_c$  gauge interactions 
are enhanced during and some time after inflation, due to the dynamics of a flat direction.
After inflation ends, the axion dynamics generates  baryon asymmetry via spontaneous baryogenesis. 
We also show that the scenario can be naturally realized when the scalar field is identified with 
the $H_u H_d$ flat direction.  The last section is devoted to  conclusions. 
In Appendix we briefly discuss the case in which the PQ symmetry is broken by
an interaction between the axion and the Ricci scalar.

\section{Axion isocurvature problem
\label{sec:1}}

The axion is a pseudo NG boson associated with the spontaneous breakdown of the PQ symmetry,
which is assumed to be anomalous under QCD~\cite{Weinberg:1977ma}. 
Suppose that the PQ symmetry is spontaneously broken at an intermediate scale $v_a$. 
Then, the axion acquires a non-zero mass through non-perturbative effects of the QCD instantons as~\cite{'tHooft:1976up, 'tHooft:1976fv}
\beq
 m_a \simeq \frac{m_u m_d}{\lmk m_u + m_d \rmk^2} \frac{m_\pi f_\pi}{f_a},
 \label{m_a at T=0}
\eeq
where 
$f_a$ ($=v_a / N_{\rm DW}$) is the axion decay constant 
and $N_{\rm DW}$ is the domain wall number. 
Here, $f_\pi$ is the pion decay constant, and $m_u$, $m_d$, and $m_\pi$ are the masses of up quark, down quark, and pion, respectively. 

The axion stays almost massless at high temperatures, and its mass gradually grows
as the temperature $T$ decreases down to the QCD scale $\LQCD$.
In a thermal plasma with temperature $T \gg \LQCD$,  the axion mass is given by ~\cite{Kim:2008hd,Wantz:2009it}
\beq
 m_a^2 (T) 
 \simeq 1.68 \times 10^{-7}\, \frac{\LQCD^4}{f_a^2} \lmk \frac{T}{\LQCD} \rmk^{-n}, 
 \label{m_a at T ne 0}
\eeq
where $n \approx 6.68$. 
When the axion mass becomes comparable to the Hubble parameter,
it starts to oscillate about the CP conserving minimum.
The energy density of the axion oscillations 
decreases in proportion to $a^{-3}$, where $a$ is the scale factor, 
and hence it is a good candidate for cold dark matter (CDM). 
Neglecting the anharmonic effect~\cite{Lyth:1991ub, Kobayashi:2013nva},
the axion abundance is given by~\cite{Bae:2008ue}
\beq
 \Omega_a h^2 
 \simeq 
 0.011 \abs{\theta_{\rm ini}}^2 
 \lmk \frac{f_a}{10^{11} \GEV} \rmk^{1.19} 
  \lmk \frac{\Lqcd}{400 \MEV} \rmk, 
  \label{axion abundance}
\eeq
where $h$ is the Hubble parameter in units of $100 \ {\rm km/s/Mpc}$ 
and $\theta_{\rm ini}$  is the initial misalignment angle. Barring fine-tuned cancellations, $\theta_{\rm ini}$ is
considered to be of order unity. The observed DM abundance $\Omega_{\rm DM} h^2 \simeq 0.12$ can be explained if the axion decay constant is given by 
\beq
 f_a \simeq 7.4 \times 10^{11} \GEV \times \abs{\theta_{\rm ini}}^{-1.68}, 
 \label{f_a for DM}
\eeq
where we have substituted $\Lqcd = 400 \MEV$. 

If the axion mass is smaller than the Hubble parameter during inflation, 
it acquires quantum fluctuations as
\beq
 \abs{\delta \theta_{\rm ini}} \equiv \frac{\delta a}{f_a}\simeq \frac{H_{{\rm inf}}}{2 \pi f_a }, 
\eeq
where $H_{\rm inf}$ is the Hubble parameter during inflation. This results in
the axion CDM isocurvature perturbations  through the dependence of $\Omega_a$ on $\theta_{\rm ini}$ (see Eq.~(\ref{axion abundance})),
\beq
 \mathcal{S}_{a \gamma} \equiv \frac{\delta \Omega_a}{\Omega_a} \simeq 2 \frac{\delta \theta_{\rm ini}}{\theta_{\rm ini}},
\eeq
where $|\delta \theta_{\rm ini}/\theta_{\rm ini}| \ll 1$ is assumed.

The cosmological data can be explained by purely adiabatic density perturbations,
and only a small admixture of isocurvature perturbations is allowed.
In fact, the $Planck$ Collaboration derived a tight upper bound on 
the fraction of the uncorrelated isocurvature perturbation
as~\cite{Ade:2015lrj}
\beq
 \frac{\mathcal{P_{SS}}(k_*)}{\mathcal{P_{RR}}(k_*) + \mathcal{P_{SS}}(k_*)} 
 \lesssim 0.038 ~~ (95\% \ {\rm C.L.}), 
\eeq
where $\mathcal{P_{RR}}$ and $\mathcal{P_{SS}}$ are the power spectra of 
the adiabatic and isocurvature perturbations, respectively, 
and $k_*$ ($=0.05 \ {\rm Mpc}^{-1}$) is the pivot scale. 
Thus we obtain an upper bound on the axion isocurvature perturbations as 
\beq
 \left\vert S_{a \gamma} \right\vert \lesssim 
 9.1 \times 10^{-6}, 
\eeq
where we have used $\mathcal{P_{RR}} \simeq 2.2 \times 10^{-9}$~\cite{Ade:2015lrj}. 
Therefore, the axion isocurvature perturbations  tightly constrain the energy scale of inflation as
\beq
 H_{\rm inf} \lesssim 
 0.94 \times 10^7 \GEV \lmk \frac{f_a}{10^{11} \GEV} \rmk^{0.405}.
 \label{isocurvature constraint}
\eeq
The constraint becomes even severer when the anharmonic effect is relevant around
$\theta_{\rm ini} \simeq \pi$~\cite{Lyth:1991ub, Kobayashi:2013nva}.
The upper bound is inconsistent with high or intermediate scale inflation models, 
such as the $R^2$-inflation model~\cite{Starobinsky:1980te}, 
chaotic inflation model~\cite{Linde:1983gd}, 
and hybrid inflation model~\cite{Linde:1993cn}. In particular, if the primordial B-mode polarization is detected
by the future CMB polarization experiment,
it would be in a strong tension with the axion CDM scenario~\cite{Higaki:2014ooa,Marsh:2014qoa,Visinelli:2014twa}.

There is one caveat in the above analysis: the axion is assumed to be massless during inflation.
In the rest of this paper, we study axion models with
a larger breaking of the PQ symmetry during inflation in order to avoid the constraint of Eq.~(\ref{isocurvature constraint}).

\section{QCD axion models with dynamically broken PQ symmetry 
\label{sec:model2}}

Now we introduce supersymmetric axion models in which the PQ symmetry is broken by a larger amount in the early Universe
either by hidden gauge interactions or by the $SU(3)_c$ strong interactions. We first consider the former case in the following, and 
we will show that the axion isocurvature perturbations  can indeed be suppressed. 
Then we investigate the dynamics of the axion after inflation and show that a right amount of baryon asymmetry can be 
generated by spontaneous baryogenesis.  In Sec.~\ref{sec:consistency}, 
we estimate the amount of axion DM and self-interacting dark radiation made of weakly-coupled hidden gauge fields.
Finally,  we introduce a more economical model 
where the PQ symmetry is broken by the stronger QCD  and a hidden flat direction is replaced by the $H_u H_d$ flat direction.

\subsection{PQ symmetry breaking by hidden gauge interactions}

Let us consider a model in which the PQ symmetry is dynamically broken by hidden $SU(N)_H$ gauge interactions 
in the early Universe.
We introduce a PQ breaking scalar field $\psi$, a singlet scalar field $\phi$,  and $N_F(N_F')$ hidden quarks and anti-quarks
$Q_H,\bar{Q}_H$  $(Q'_H,\bar{Q}'_H)$ in the fundamental and anti-fundamental representation of $SU(N)_H$,  with (without) PQ charges.
The charge assignment is shown in Table~\ref{table1}.%
\footnote{
In general, the high quality of the PQ symmetry is a puzzle 
because global symmetries are considered to be broken 
in nature~\cite{Georgi:1981pu,Lazarides:1985bj,Dine:1986bg,Kamionkowski:1992mf, Holman:1992us,Barr:1992qq,Dobrescu:1996jp}. 
This puzzle is neither worsened or improved compared with the ordinary field-theoretic QCD axion
models, because the extra PQ breaking effects we introduced disappear in the present vacuum.  The required PQ breaking scale in our model
is of order $10^{13}$ GeV. This implies that  $Z_N$  symmetry with $N \geq 17$ is necessary
to suppress Planck-suppressed  PQ breaking operators, if we are to explain
the lightness of the QCD axion by a single discrete symmetry~\cite{Dine:1992ya}. 
There are also some other mechanisms to explain the origin of the PQ symmetry (see Refs.~\cite{Svrcek:2006yi, Barr:2001vh, Izawa:2002qk, Izawa:2004bi, Harigaya:2015soa, Kawasaki:2015lea}). 
}
For simplicity, we assume that the fields $Q_H$ and $\bar{Q}_H$ are also charged under $SU(3)_c$ in
the fundamental and anti-fundamental representation, respectively so that they play a role of the heavy PQ quarks to induce
the color anomaly term for the axion~\cite{Kim:1979if}.\footnote{Instead, one may introduce heavy PQ quarks separately, if
$Q_H$ and $\bar{Q}_H$ are singlet under $SU(3)_c$.}

\begin{table}\begin{center}
{\renewcommand\arraystretch{1.5}
\begin{tabular}{|c|p{1.4cm}|p{1.4cm}|p{1.4cm}|p{1.4cm}|p{1.4cm}|p{1.4cm}|}
  \hline
    & \hfil $\psi$ \hfil & \hfil $Q_H$ \hfil & \hfil $\bar{Q}_H$ \hfil & \hfil $\phi$ \hfil & \hfil $Q'_H$ \hfil & \hfil $\bar{Q}'_H$ \hfil  \\
  \hline
  \hfil $SU(3)_c$ \hfil & \hfil {\bf 1} \hfil & \hfil {\bf 3} \hfil & \hfil ${\bf 3}^*$ \hfil & \hfil {\bf 1} \hfil & \hfil {\bf 1} \hfil & \hfil {\bf 1} \hfil   \\
  \hline
  \hfil $SU(N)_{\rm H}$ \hfil & \hfil {\bf 1} \hfil & \hfil {\bf N} \hfil & \hfil ${\bf N}^*$ \hfil & \hfil {\bf 1} \hfil & \hfil {\bf N} \hfil & \hfil ${\bf N}^*$ \hfil  \\
  \hline
  \hfil $U(1)_{\rm PQ}$ \hfil & \hfil $1$ \hfil & \hfil $-1/2$ \hfil & \hfil $-1/2$ \hfil & \hfil $0$ \hfil & \hfil 0 \hfil & \hfil 0 \hfil \\
\hline
\end{tabular}
}
\end{center}
\caption{Charge assignment for matter chiral superfields.
\label{table1}}
\end{table}

We consider the superpotential of 
\beq
 W =  y_{ij} \,\psi\, Q_H^i \bar{Q}_H^j + 
 y'_{kl}\,\phi\, Q_H^{'k} \bar{Q}_H^{'l} +W_{\rm PQ} 
 + W_\phi (\phi), 
 \label{superpotential1}
\eeq
where $i$, $j$, $k$, and $l$ are flavor indices 
and we omit $SU(N)_H$ and $SU(3)_c$ indices which are contracted in a gauge invariant way.%
\footnote{
In Ref.~\cite{Barr:2014vva}, they investigated a similar non-SUSY model. 
They have considered a scenario that the PQ symmetry is largely broken after inflation due to the strong dynamics of hidden gauge interaction. 
They assume $N_F = 1$ so that domain walls decay soon after the phase transition of the hidden gauge symmetry. 
}
The superpotential of $W_{\rm PQ}$ represents interaction terms for $\psi$ to develop a nonzero 
VEV which spontaneously breaks the PQ symmetry.
We assume that the axion decay constant (the PQ breaking scale) remains constant throughout the history of the Universe. 
While we have introduced only a single PQ field $\psi$  for simplicity, we may introduce
another PQ scalar $\bar\psi$ to write down a superpotential like
$W_{\rm PQ} \supset \kappa S ( \psi \bar{\psi} - v_a^2/4)$, 
where $\kappa$ is a parameter and $S$  is a singlet chiral superfield.
Assuming an (approximate) exchange symmetry between $\psi$ and $\bar\psi$, they are
stabilized at $|\psi| = |\bar \psi| = v_a/2$. 
The axion is given by a combination of the phases of $\psi$ and $\bar \psi$.
As long as the inflation scale is lower than $v_a$, the PQ breaking scale remains almost constant in this case.  

Let us first focus on the axion in the present Universe, assuming that the hidden $SU(N)_H$ is weakly coupled and
 $\phi$ is stabilized at the origin (or its VEV is sufficiently small). 
 For instance, in the non-SUSY limit where the SUSY particles are integrated out, the hidden $SU(N)_H$
 is asymptotic non-free if $N_f' > 11 N/4$. 
  Then, the phase of $\psi$ (or a combination of the phases of $\psi$ and $\bar \psi$ in the
above example) becomes the axion, and its anomalous coupling to gluon
fields is induced by the diagram with $Q_H$ and $\bar Q_H$ running in the loop.
The domain wall number $N_{\rm DW}$ is equal to $N N_F$ in this model. The axion is stabilized at the CP conserving
minimum, solving the strong CP problem. Note that the PQ symmetry is anomalous under $SU(N)_H$,
which however does not affect the axion mass as long as $SU(N)_H$ is weakly coupled.

The situation is different and more complicated if $\phi$ develops a large VEV in the early Universe. 
Before going into details, let us briefly outline a rough sketch of our scenario. 
First, as we shall see below, the $U(1)_R$ symmetry is spontaneously broken by a non-zero VEV of $\phi$,
and the phase component $\phi$ becomes a pseudo NG boson called the $R$-axion.\footnote{
Depending on the $R$ charge of $\psi$, the $R$-axion may be composed of a combination of the phases of $\phi$ and $\psi$.
This however does not change our arguments. }
The $U(1)_R$ symmetry is not an exact symmetry of nature, and it is explicitly broken by a constant term in the
superpotential, $W \supset W_0 = m_{3/2} \Mpl^2$, where $m_{3/2}$ is the gravitino mass. It is possible that the inflaton sector
also  does not respect the $U(1)_R$ symmetry, in which case the $R$-axion can be so heavy that it does not acquire 
sizable quantum fluctuations at superhorizon scales.  In general, the  R-axion mass is given by~\cite{Bagger:1994hh} 
\beq
 m_R^2 = \frac{8}{f_R^2} \frac{W}{\Mpl^2} \abs{F_i \phi_i - 3 W}, 
\eeq
where the R-axion coupling $f_R$ is given by 
\beq
 f_R = \sqrt{2} R_i \phi_i, 
\eeq
where $R_i$ is the $R$ charge of field $\phi_i$ 
and $\phi_i$ represents a field which spontaneously breaks the $U(1)_R$ symmetry.
We assume that the $U(1)_R$ symmetry is explicitly broken in the inflaton sector  so that the R-axion mass 
is comparable to (or heavier than) the Hubble parameter during inflation. This is the case if
 $H^2 \sim \abs{F_i}^2 \sim \abs{W}^2 / \phi_i^2$. 
In other words,  the mass of the phase component of  $\phi$ is of order the Hubble parameter 
due to the Hubble-induced $A$-term. 
Secondly,
the hidden quarks $Q_H'$ and $\bar Q_H'$ acquire a heavy mass through its coupling to $\phi$, 
and they are decoupled from the low-energy physics. As a result, the running of the gauge coupling $g_H$ of $SU(N)_H$ 
is modified, so that it becomes confined during inflation if the confinement scale is higher than $H_{\rm inf}$.
 This implies that the PQ symmetry is broken badly by non-perturbative effects
 of $SU(N)_H$. Thus, the axion mass receives an extra contribution during inflation,
 suppressing the quantum fluctuations. The axion dynamics just after inflation is involved as we shall study in detail later in this section.
 Some time after inflation the potential minimum of $\phi$ shifts to the origin, and the $SU(N)_H$
 becomes weakly coupled. The axion becomes almost massless and it remains so until the QCD phase transition.

Now let us discuss the dynamics of $\phi$ in a greater detail. We assume that $\phi$ is a flat direction, namely, 
the potential of $\phi$ is flat in the SUSY limit at the renormalizable level, which can be ensured by assigning
a certain $R$ charge or a discrete symmetry. Such flat directions can be lifted by soft SUSY breaking terms
as well as non-renormalizable terms.
To be concrete we consider the following superpotential for $\phi$: 
\beq
 W_\phi (\phi) = \lambda \frac{\phi^4}{4 \Mpl},
 \label{superpotential of phi}
\eeq
where we assign $R$-charges as $R(\phi) = 1/2$ and $R(Q_H' \bar Q_H') = 3/2$,
and $\Mpl \simeq 2.4 \times 10^{18}$\,GeV is the reduced Planck mass.
During inflation, 
the potential of $\phi$ is given by
\beq
 V(\phi) = m_\phi^2 \abs{\phi}^2 
 - c_H H^2 \abs{\phi}^2 
 - \lmk a_H \lambda H \frac{\phi^4}{4 \Mpl} + {\rm c.c.} \rmk
 + \lambda^2 \frac{\abs{\phi}^6}{\Mpl^2}, 
\label{V_phi}
\eeq
where $m_\phi$ is the soft  SUSY breaking mass of $\phi$ 
and 
the second term in the RHS is the Hubble-induced mass term~\cite{Dine:1995kz}. 
The term in the parenthesis is the Hubble-induced $A$-terms, 
which generically arises when the inflaton breaks the $U(1)_R$ symmetry. 
Hereafter, we set $c_H = a_H = 1$ for simplicity. 

During inflation,
since the flat direction has a mass of order the Hubble parameter, 
the $\phi$ is stabilized at its potential minimum: 
\beq 
 \la \abs{\phi} \ra (t)  \simeq \lmk \frac{H(t) \Mpl}{\sqrt{3} \lambda} \rmk^{1/2}. 
 \label{VEV}
\eeq
The phase direction of $\phi$ also obtains a mass of order the Hubble parameter 
due to the Hubble-induced $A$-terms. Therefore,
it stays at $\theta = 0$ and it does not acquire any sizable quantum fluctuations
at large scales.

After inflation ends,  the inflaton (not $\phi$) starts to oscillate around the potential minimum 
and the energy density of the Universe is dominated by the inflaton oscillation. 
During the inflaton oscillation dominated era, there is a dilute plasma with
temperature, 
\beq
 T \simeq \lmk \frac{36 H(t) \Gamma_I \Mpl^2}{g_*(T) \pi^2 } \rmk^{1/4} 
 \quad  \propto a^{-3/8}, 
 \label{T before reheating}
\eeq
where $H(t) = 2/(3t)$, and $\Gamma_I$ is the inflaton decay rate.
Note that, if the plasma temperature is higher than $\sim y v_a$, 
 the hidden $SU(N)_H$ gauge fields as well as the hidden quarks are
thermalized because $Q_H$ and $\bar Q_H$ are charged under both $SU(3)_c$ and
$SU(N)_H$.\footnote{
The PQ symmetry remains broken if $y \ll 1$.
}  At a sufficiently high temperature, 
the effective relativistic degrees of freedom $g_*$ is given as 
$g_* = 228.75$ in the MSSM. Hereafter, we use $g_* \simeq 200$ as a reference value.\footnote{
To be precise, we should include the degrees of freedom of hidden fields 
in $g_*$, which however does not affect our main results qualitatively.
In Sec.~\ref{sec:Higgs},  many SM particles are decoupled in the thermal plasma due to 
a large VEV of the $H_u H_d$ flat direction, 
so that $g_*$ is smaller.
}  
The reheating temperature is  given by 
\beq
 T_{\rm RH} 
 &\simeq& \lmk \frac{90}{g_*(T_{\rm RH}) \pi^2} \rmk^{1/4} \sqrt{\Gamma_I \Mpl} \\
 &\simeq& 1.0 \times 10^{13} \GEV 
 \lmk \frac{\Gamma_I}{2 \times 10^8 \GEV} \rmk^{1/2}. 
\eeq
In a finite temperature, 
the following thermal log potential is induced via two-loop effects:~\cite{Asaka:2000nb, Anisimov:2000wx} 
\beq
 V_T (\phi) \simeq 
 c_T \alpha_H^2 T^4 \log \lmk \frac{ \abs{\phi}^2 }{T^2} \rmk, 
\eeq
for $\phi \gg T/g_H$, 
where $\alpha_H = g_H^2 / 4 \pi$ is the fine-structure constant of $SU(N)_H$ 
and $c_T$ ($> 0$) is an $\mathcal{O}(1)$ constant determined by the beta function of $SU(N)_H$ coupling. 

The flat direction $\phi$ starts to oscillate  at $t = t_{\rm osc}^\phi$ when the Hubble parameter
becomes comparable to the soft mass  or the curvature of thermal potential.
The Hubble parameter at the commencement of oscillations of $\phi$ is given by
\beq
 H^{\phi}_{\rm osc} \simeq {\rm Max} \lkk m_\phi , \ \lmk \frac{72}{5 g_* \pi^2} \rmk^{1/2} \frac{c_T \alpha_H^2 T_{\rm RH}^2 \Mpl}{\la \abs{\phi} \ra^2 (t^{\phi}_{\rm osc}) } \rkk. 
 \label{H_osc^phi}
\eeq
This implies that $\phi$ starts to oscillate around the origin 
before reheating completes, i.e., $T^{\phi}_{\rm osc} > T_{\rm RH}$. 
In our scenario, 
we find that 
$H^{\phi}_{\rm osc}$ is  determined by the second term for the parameters of our interest. 
Substituting $\mathcal{O}(1)$ parameters, 
we obtain a typical value of $H^{\phi}_{\rm osc}$ as 
\beq
 H^{\phi}_{\rm osc} \simeq 
 1.4 \times 10^{9} \GEV 
 \lmk \frac{T_{\rm RH}}{10^{13} \GEV} \rmk^{2} 
 \lmk \frac{\la \abs{\phi} \ra^2 (t^{\phi}_{\rm osc}) }{5 \times 10^{15} \GEV} \rmk^{-2}, 
\eeq
where we have adopted $\alpha_H \simeq 1/25$ as a reference value.
This implies that $T^{\phi}_{\rm osc} \simeq 3.3 \times 10^{13} \GEV$ 
for the above reference parameters.

\subsection{Suppression of the axion CDM isocurvature perturbations}

When 
the flat direction $\phi$ obtains a large VEV as  Eq.~(\ref{VEV}), 
its $F$-component is given by
\beq
 \la F_\phi \ra \simeq \lmk \frac{H^3 (t) \Mpl}{3^{3/2} \lambda} \rmk^{1/2}, 
 \label{F_phi}
\eeq
from Eq.~(\ref{superpotential of phi}). 
This implies that the $SU(N)_H$ gaugino obtains a soft mass 
via the gauge mediated SUSY breaking effect: 
\beq
 m_\lambda 
 &\simeq& \frac{N'_F g_H^2}{16 \pi^2} \frac{F_\phi}{\la \phi \ra} \\
 &\simeq& N'_F \frac{g_H^2}{16 \sqrt{3} \pi^2} H(t). 
\eeq
We may write the Lagrangian of the $SU(N)_H$ gauge field as 
\beq
 \mathcal{L} = \int \dd^2 \theta \frac{\tau_H}{2} W^\alpha W_\alpha + {\rm H.c.}, 
\eeq
where 
\beq
 \tau_H = \frac{1}{2 g_H^2} - \frac{i  \theta_H  }{16 \pi^2} - \theta^2 \frac{m_\lambda}{g_H^2}, 
 \label{tauH}
\eeq
is a chiral spurion superfield that contains the gauge coupling $g_H$, vacuum angle $\theta_H$,  
and gaugino mass $m_\lambda$~\cite{Giudice:1997ni, ArkaniHamed:1998kj}.
The axion appears in the gauge kinetic function, and it can be taken into account by the replacement of
\beq
\theta_H \to \theta_H - 3 N_F \frac{a}{ v_a} = \theta_H - \frac{3}{N} \frac{a}{ f_a},
\label{repa}
\eeq
where we have used $f_a = v_a / N_{\rm DW } = v_a / (N N_F)$ in the second equality.

\begin{figure}[t]
\centering 
\includegraphics[width=.75\textwidth, bb=0 0 842 595
]{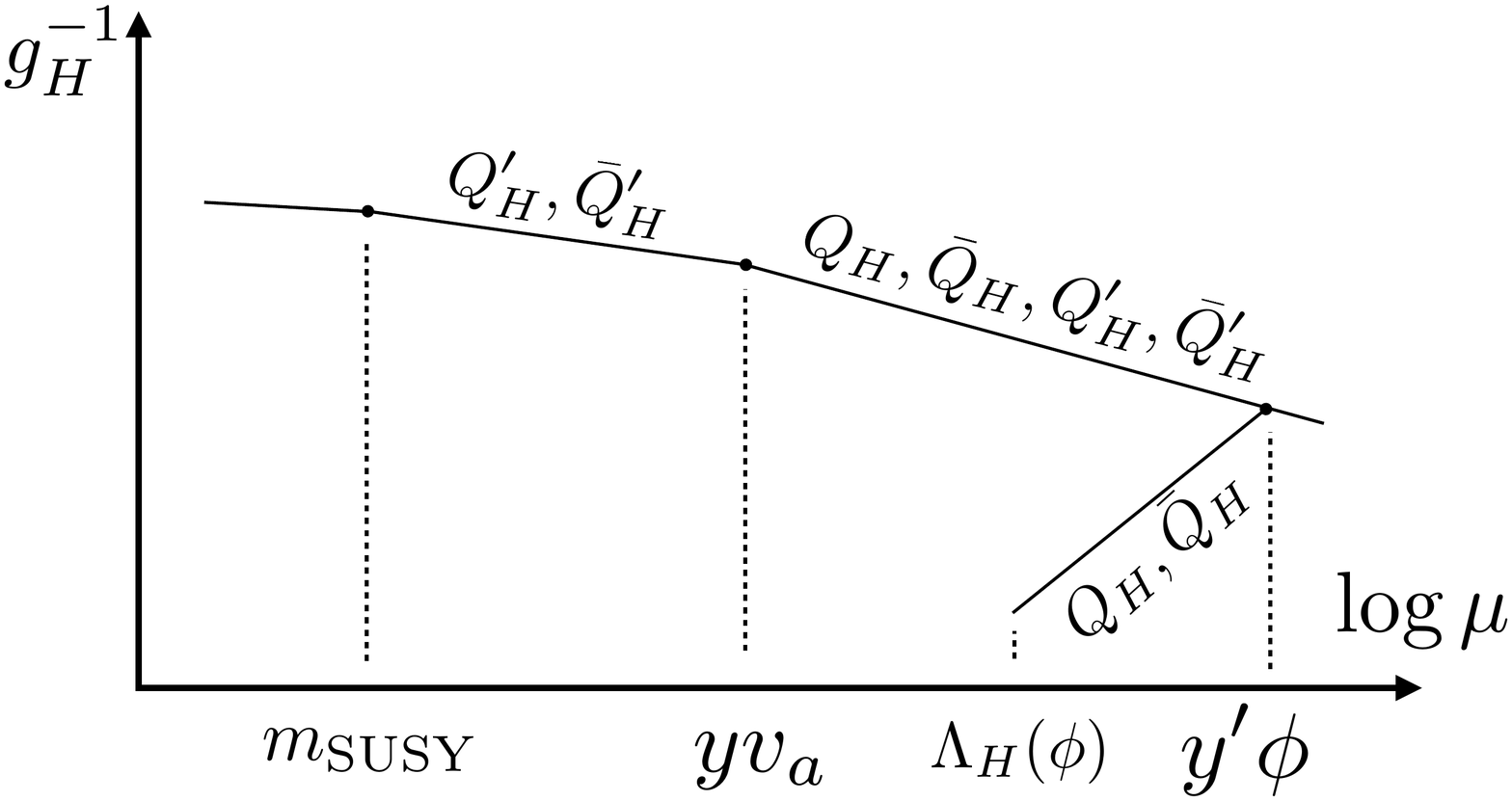} 
\caption{\small
Renormalization group flow of the $SU(N)_H$ gauge coupling $g_H$. 
It is weakly coupled at present for sufficiently large $N_f'$, 
while it is strongly coupled at the energy scale of $\Lambda_H (\phi)$ 
if  $\phi$ develops a large VEV in the early Universe. 
}
  \label{fig:fig1}
\end{figure}

Next, we calculate the confinement scale of $SU(N)_H$ in order to determine the axion mass.
First, let us consider the present era, where 
the VEV of $\phi$ is absent, $\la \phi \ra = 0$. 
In this case, 
the $SU(N)_H$ gauge theory contains 
$N'_F$ flavors of $Q'_H$ and $\bar{Q'}_H$. 
For a sufficiently large $N_F'$, $SU(N)_H$ remains weakly coupled at present.
Above the soft SUSY breaking scale $m_{\rm SUSY}$, 
the running gauge coupling $g_H$ is given by
\beq
 \frac{1}{g_H^2 (\mu)} - \frac{1}{g_H^2 (m_{\rm SUSY})} 
 = - \frac{3N - N'_F}{8 \pi^2} \ln \lmk \frac{m_{\rm SUSY}}{\mu} \rmk, 
\eeq
where $\mu$ ($m_{\rm SUSY} < \mu < y v_a$) is the renormalization scale. 
Next, suppose that the field $\phi$ has a large VEV as Eq.~(\ref{VEV}) in the early Universe. 
The fields $Q'_H$ and $\bar{Q}'_H$ then obtain heavy masses 
due to the nonzero VEV of $\phi$, which modify  the running of the gauge coupling (see Fig.~\ref{fig:fig1}). 
The $SU(N)_H$  confines during inflation if the confinement scale $\Lambda_H$ is higher than the Hubble parameter
during inflation, $H_{\rm inf}$.
The confinement scale $\Lambda_H$ depends on $\phi$ as 
\beq
 \Lambda_H^{3N} (\phi) = 
 \det \lmk y' \phi \rmk m_{\rm SUSY}^{3N - N'_F} e^{-8 \pi^2 / g_H^2(m_{\rm SUSY})}, 
\eeq
because the renormalization group equation changes at the energy scale of $y' \phi$. 
For later convenience, we define $\tilde{\Lambda}_H$ by 
\beq
 \tilde{\Lambda}_H^{3N} (\phi)
 \equiv 
 \det \lmk y' \phi \rmk m_{\rm SUSY}^{3N - N'_F} e^{-16 \pi^2 \tau_H (m_{\rm SUSY})}, 
 \label{tilde_Lambda}
\eeq
where $ \tau_H$ is given by Eq.~(\ref{tauH}) with the replacement of Eq.~(\ref{repa}).

At the confinement scale  $\Lambda_H$,  the gaugino condensation gives rise to 
the effective superpotential given by~\cite{Affleck:1983mk}
\beq
 W_{\rm eff} = N \tilde{\Lambda}_H^3 (\phi). 
\label{W_eff}
\eeq
Therefore, 
the field $a$ acquires an effective potential 
such as 
\beq
 V &=& 
 - \int \dd^2 \theta W_{\rm eff} + {\rm H.c.} 
 \\
 &=& 
 \frac{32 \pi^2}{g_H^2} m_\lambda \Lambda_H^3 
 \cos \lmk \frac{ \theta_H}{N} - \frac{3 a}{N^2 f_a} \rmk 
 + \dots, 
\eeq
where we use Eqs.~(\ref{F_phi}), (\ref{tilde_Lambda}), and (\ref{W_eff}). 
This gives the field $a$ an effective mass of 
\beq
 m_a^2 (\phi)
 &=& c_m \frac{H_{\rm inf} \Lambda_H^3 (\phi)}{f_a^2} \\ 
 &\simeq& 
 \lmk 1.0 \times 10^{12} \GEV \rmk^2 
 \times c_m 
 \lmk \frac{H_{\rm inf}}{10^{11} \GEV} \rmk \nonumber\\
 &&\qquad \times 
 \lmk \frac{f_a}{5 \times 10^{13} \GEV} \rmk^{-2} 
 \lmk \frac{\Lambda_H (\phi)}{4 \times 10^{13} \GEV} \rmk^3, 
 \label{m_a}
\eeq
where 
\beq
 c_m \equiv  \frac{6 \sqrt{3} N'_F }{N^4}. 
\eeq
When $m_a (\la \abs{\phi} \ra_{\rm inf})> H_{\rm inf}$ is satisfied, 
the field $a$ does not acquire  quantum fluctuations during inflation,
thus avoiding the isocurvature constraint.

\subsection{Spontaneous baryogenesis via axion dynamics}

Now let us consider the axion dynamics after inflation. We shall show that spontaneous baryogenesis
by the QCD axion works for certain parameters as one application of the strongly broken PQ symmetry.

We assume that $SU(N)_H$ is deconfined just after inflation. 
This can be achieved if the
inflaton decays into the $SU(N)_H$ gauge field as well as standard model particles 
and the maximum temperature of the Universe (or the hidden sector, if it is decoupled
from the standard model sector) is larger than 
the confiment scale $\Lambda_H$: 
\beq
 T_{\rm max} \gtrsim  \Lambda_H (\la \abs{\phi} \ra_{\rm inf}). 
\eeq
Here, 
the maximal temperature of the Universe after inflation 
is given by%
\footnote{
At such a high temperature, axions can be in the thermal equilibrium via interactions such as $g + g \leftrightarrow a + g$, 
where $g$ represents gluon~\cite{Turner:1986tb, Masso:2002np}. 
Then the $SU(N)_H$ hidden gauge fields are also in the thermal equilibrium 
via similar interactions 
as $g_H + g_H \leftrightarrow a + g_H$, where $g_H$ represents hidden gluon. 
Also, if the temperature is higher than $y v_a$, $Q_H$ and $\bar Q_H$ are thermalized.
Thus $SU(N)_H$ is in the thermal equilibrium 
and is deconfined after inflation 
even if the inflaton decays only into the SM particles. 
}
\beq
 T_{\rm max} 
 &\simeq& 
 \lkk \frac{60}{g_* \pi^2} \lmk \frac{3}{8} \rmk^{8/5} \Gamma_I H_{\rm inf} \Mpl^2 \rkk^{1/4}, \\
 &\simeq& 
 2.9 \times 10^{13} \GEV 
 \lmk \frac{\Gamma_I}{0.2 \times 10^9 \GEV} \rmk^{1/4} 
 \lmk \frac{H_{\rm inf}}{10^{11} \GEV} \rmk^{1/4}. 
\eeq
We also require the following condition 
to avoid the restoration of PQ symmetry by thermal effects: 
\beq 
 f_a \gtrsim T_{\rm max}. 
\eeq

Now, let us explain the dynamics of the axion and $\phi$ in our scenario. 
During inflation, the axion acquires the effective mass of Eq.~(\ref{m_a}), 
which is assumed to be larger than the Hubble parameter to suppress the isocurvature perturbation. 
After inflation ends, 
the temperature of the Universe soon reaches the maximal temperature $T_{\rm max}$,
and then, the axion becomes massless 
because of $T_{\rm max} \gtrsim \Lambda_H (\la \abs{\phi} \ra_{\rm inf})$.
When the temperature decreases to $T = T^{a}_{\rm osc} \simeq \Lambda_H (\phi)$, 
the axion again acquires an effective mass due to instanton effects 
of $SU(N)_H$ gauge theory. As we shall see shortly,  the minimum could be different from that during inflation, 
in which case the axion starts to oscillate about the new minimum. Here, the minimum of the axion potential 
is given by $\theta_H f_a$, 
which is determined as
\beq
 \theta_H = \theta_{H,0} + \arg \lkk \det \lmk y \la \phi \ra \rmk \rkk 
 + \arg \lkk \det \lmk y' \la \phi \ra \rmk \rkk 
 + N \arg \lkk m_{\lambda, 0} \rkk, 
\label{theta_H}
\eeq
where $\theta_{H,0}$ is the bare theta parameter 
and 
$m_{\lambda, 0}$ is the mass of $SU(N)_H$ gaugino. 
Since the minimum of the axion field value is related to $\theta_H$, 
it also depends on 
the phase of $U(1)_R$-symmetry breaking term 
through $\arg \lkk m_{\lambda,0} \rkk$. 
As the $U(1)_R$ symmetry is assumed to be largely broken in the inflaton sector, 
the phase of $U(1)_R$-symmetry breaking term generically changes after inflation, 
which  implies that 
the minimum of the axion potential is shifted by an amount of $\Delta a \simeq \mathcal{O}(1) f_a$. 
Therefore,  the
axion starts to oscillate around the new minimum at $T = T^{a}_{\rm osc} \simeq \Lambda_H(\phi)$. 
This axion dynamics breaks the CPT invariance, which enables the spontaneous baryogenesis~\cite{Cohen:1987vi,Dine:1990fj,Cohen:1991iu,Kusenko:2014uta,Daido:2015gqa}.

Let us focus on the dynamics of axion at the temperature around $T = T^{a}_{\rm osc}$ 
and consider the spontaneous baryogenesis based on Ref.~\cite{Kusenko:2014uta}. 
We assume that 
the axion couples to the electroweak $SU(2)$ gauge fields as 
\beq
 \mathcal{L} \supset \frac{g_2^2}{32 \pi^2} \frac{a}{f_a} F \tilde{F}, 
 \label{aFF}
\eeq
where $g_2$ is the $SU(2)$ gauge coupling. 
Using the anomaly equation of the $SU(2)$ gauge theory, 
we obtain the following derivative couplings between 
axion and SM particles: 
\beq
 \mathcal{L} 
 &=& - c^a \frac{a(t)}{f_a} \del_\mu j^\mu_{B+L} \\
 &=& c^a \frac{\del_0 a}{f_a} j^0_{B+L} + \dots, 
\label{anomaly eq}
\eeq
where $c_i^a = 1/ 3$ and $j^\mu_{B+L}$ is the $B+L$ current. Note that this deformation is valid if
the $SU(2)_L$ sphaleron effect is sufficiently efficient~\cite{Daido:2015gqa}.
The sphaleron rate is given by $\Gamma_{\rm sphaleron} \sim 25 \,\alpha_2^5 T^4$ per 
unit time and volume~\cite{Bodeker:1999gx}. 
Taking the thermal volume, $1/T^3$, and comparing it to the Hubble rate, 
the sphaleron decouples above the temperature of order $10^{13} \GEV$. 
This is of the same order with $T_{\rm osc}^\phi$, 
so that our scenario is marginally consistent. 
Note that the above sphaleron decoupling temeprature is calculated 
by assuming $\alpha_2 \sim 1/25$. 
Since we introduce some $SU(3)_c$ charged fields, such as $Q_H$ and $\bar{Q}_H$, 
the unified gauge coupling constant may be larger than $1/25$.\footnote{
One needs to add extra matter fields in order to form complete multiplets under $SU(5)$,
which would keep the gauge coupling unification intact. 
}
This implies that the $SU(2)_L$ coupling constant may also be larger than $1/25$ at a high temperature. 
Therefore, the sphaleron decoupling temperature may be higher than $10^{13} \GEV$ in our model 
and we can safely use Eq.~(\ref{anomaly eq}). 

One needs baryon or lepton number violating interactions for successful spontaneous baryogenesis.
We shall focus on the lepton number violating operator mediated by right-handed neutrinos, and so, 
let us focus on the lepton asymmetry in the following.
From Eq.~(\ref{anomaly eq}), the axion dynamics leads to an effective chemical potential for lepton asymmetry as 
\beq
 \mu_{\rm eff} = 
 c^a \frac{\del_0 a}{f_a}. 
 \label{mu}
\eeq
At the temperature of $T = T^{a}_{\rm osc}$, 
the axion starts to oscillate around the minimum 
and the effective chemical potential is given by $\mu_{\rm eff} \simeq c_a H(T^{a}_{\rm osc}) \Delta a / f_a$. 
The nonzero effective chemical potential 
results in the following equilibrium number density: 
\beq
 n_L^{\rm eq} = \frac{1}{6} \mu_{\rm eff} T^2.
\eeq
Note that the equilibrium number density is realized {\it only if} the baryon/lepton number is
explicitly broken and its rate is sufficiently rapid.

Now we introduce right-handed neutrinos for the seesaw mechanism~\cite{seesaw} 
and assume that their masses are close to $10^{15}$\,GeV.
Thermal leptogenesis~\cite{Fukugita:1986hr} does not work for such heavy right-handed neutrinos.
Nevertheless, they provide lepton number violating processes, $l l \leftrightarrow H H$ and $l H \leftrightarrow \bar{l} \bar{H}$, 
where $l$ and $H$  represent left-handed lepton and higgs multiplets, respectively. 
The effective lepton number violating rate is roughly given by 
\beq
\Gamma_{L} 
\sim
\frac{\bar{m}^2 T^3}{16 \pi v^4_{\rm ew}}, 
\eeq
where $\bar{m}^2$ is the sum of the left-handed neutrino mass squared 
and is assumed to be of order the atmospheric neutrino mass squared difference, $\Delta m_{\rm atm}^2 \simeq 2.4 \times 10^{-3} \EV^2$~\cite{Agashe:2014kda}. 
This lepton violating interactions lead to a nonzero lepton asymmetry in the Universe
in the presence of the chemical potential of Eq.~(\ref{mu}).
The resulting lepton asymmetry just after $T = T^{a}_{\rm osc}$ is thus calculated as 
\beq
 \left. n_L \right\vert_{T=T^{a}_{\rm osc}} 
 &\simeq& 
 \frac{1}{H^{a}_{\rm osc}} \Gamma_L  n^{\rm eq}_L.
\label{n_L}
\eeq

After the axion starts to oscillate, 
the generated lepton asymmetry is partially washed out 
due to the inverse processes. 
This is described by the following Boltzmann equation:~\cite{Kusenko:2014lra} 
\beq
 \frac{\dd a^3 n_L (t)}{\dd t} 
 \simeq - \Gamma_L a^3 n_L(t). 
\eeq
Using Eq.~(\ref{T before reheating}) before reheating 
and $T^4 = 90 H^2 \Mpl^2 / (\pi^2 g'_*)$ after reheating, 
where $g'_*$ is the effective number of degrees of freedom for number density, 
and matching the solutions at $T = T_{\rm RH}$, 
we obtain 
\beq
 a^3 n_L (T \to 0) 
 &\equiv& 
  a^3 n_L (T = T^{a}_{\rm osc}) \Delta_w, 
\eeq
where 
\beq
 \log \Delta_w \simeq 
 -0.7 \lmk \frac{T_{\rm RH}}{10^{13} \GEV} \rmk. 
\eeq
This implies that 
the reheating temperature cannot be much larger than $10^{13} \GEV$ 
to avoid the washout effect due to the inverse processes.

Since $T = T^{a}_{\rm osc}$ occurs before the reheating completes, 
the baryon asymmetry is calculated as 
\beq
 Y_b 
 \equiv \frac{n_b}{s} 
 &\simeq& -\frac{8 n_f+4n_H}{22n_f+13 n_H} T_{\rm RH} 
 \left. \frac{3 n_L}{4 \rho_{\rm tot}} \right\vert_{\rm RH} \\
 &\simeq& 
 1.6 \times 10^{-10} \Delta_w 
 \lmk \frac{T_{\rm RH}}{10^{13} \GEV} \rmk^{7/2} 
 \lmk \frac{H^{a}_{\rm osc}}{1.4 \times 10^{9} \GEV} \rmk^{-3/4}, 
 \label{result}
\eeq
where $n_f = 3$ and $n_H =2$ in the MSSM. 
Here, we implicitly assume that 
$\phi$ and axion oscillations never dominate the Universe, 
on which we shall comment shortly.
We find that the observed amount of baryon asymmetry $Y_b^{\rm obs} \simeq 8.6 \times 10^{-11}$ 
can be explained by the above mechanism. 
Since $H^{a}_{\rm osc} \sim (T^{a}_{\rm osc})^4 T_{\rm RH}^{-2} \Mpl^{-1}$ 
and $T^{a}_{\rm osc} \gtrsim T_{\rm RH}$, 
there is a lower bound on the reheating temperature 
to explain the observed amount of baryon asymmetry: 
\beq
 T_{\rm RH}^{\rm min} \simeq 5.0 \times 10^{12} \GEV. 
\eeq
This implies that  the confinement scale $\Lambda_H$ during inflation should be
higher than $5.0 \times 10^{12} \GEV$. 

Our scenario requires the Hubble parameter of inflation to be
of order $10^{11} \GEV$.  This is because it should be smaller than the mass of the axion during inflation,
while the successful baryogenesis  requires a high reheating temperature.
The energy scale of $10^{11} \GEV$ predicts the tensor-to-scalar ratio of order $r = \mathcal{O} (10^{-7})$,
which is too small to be detected in the future CMB polarization experiments.

Finally let us study the dynamics of $\phi$ to see if
 the above scenario works successfully.
The field $\phi$ starts to oscillate around the origin of the potential 
at $H = H_{\rm osc}^\phi$, 
where $H_{\rm osc}^\phi$ is given by Eq.~(\ref{H_osc^phi}). 
After the field $\phi$ starts to oscillate, 
$\Lambda_H (\phi)$ becomes much smaller than the temperature of the Universe, 
so that the axion becomes massless again.
So far we have assumed that
\beq
 T^{a}_{\rm osc} \gtrsim T^{\phi}_{\rm osc}, 
\eeq
holds so that axion starts to oscillate before $\phi$ starts to oscillate. 
Otherwise the dynamical scale becomes much smaller than temperature 
before axion starts to oscillate to generate the baryon asymmetry. 
Here, let us check that the energy density of $\phi$ oscillation never dominate that of the Universe. 
It starts to oscillate around the origin of the potential 
by the thermal log potential at $H = H^{\phi}_{\rm osc}$, 
in which case 
it is known that 
Q-balls form after the oscillation~\cite{Coleman, KuSh, EnMc, KK1, KK2, KK3}. 
The energy density of the flat direction 
is converted to that of Q-balls. 
However, in the case of our interest,
a typical charge of Q-balls is so small that 
they completely evaporate via interactions with thermal plasma~\cite{Laine:1998rg, Banerjee:2000mb}. 
Thus, they evaporate soon and dissipate into thermal plasma, 
and their energy density never dominates the Universe.

\subsection{Axion dark matter and self-interacting dark radiation
\label{sec:consistency}}

At $T \simeq \Lambda_{\rm QCD}$, 
the axion acquires an effective mass through the non-perturbative effect 
of the $SU(3)_c$ gauge theory.
The minimum of the axion potential induced by the QCD instantons
is generally different from that by the $SU(N)_H$ instantons.
Thus, the axion again starts to oscillate around its minimum at $T  \simeq \Lambda_{\rm QCD}$ 
and we can explain the observed DM abundance by the axion oscillations
as Eq.~(\ref{axion abundance}).

Let us mention an interesting prediction of our scenario. 
To solve the strong CP problem, 
the $SU(N)_H$ gauge theory should not be confined 
at present. 
Therefore, our model predicts that 
there are at least massless hidden gauge bosons, and some of the hidden quarks $Q'_H$ and $\bar{Q}'_H$
may also remain sufficiently light.%
\footnote{
The $SU(N)_H$ can be asymptotic non-free at present if there are many massless hidden quarks.
If $\phi$ develops a small VEV, hidden quarks $Q'_H$, and $\bar{Q}'_H$ acquire a light mass and
some of them may be decoupled at present. Alternatively, if the $U(1)_R$ symmetry is broken
down to a discrete $R$ symmetry, some of them acquire a non-zero mass 
depending on the $R$-charge assignment.
The scalar components have a mass of order $m_{\rm SUSY}$
and they do not contribute to dark radiation.
}
Since those light hidden particles are in the thermal equilibrium
just after inflation ends, 
they contribute to the energy density of the Universe 
as dark radiation~\cite{Nakayama:2010vs, Weinberg:2013kea, Jeong:2013eza,Kawasaki:2015ofa}. 
Their abundance is commonly expressed by the effective neutrino number 
and is calculated as~\cite{Nakayama:2010vs}
\beq
 \Delta N_{\rm eff} &=& 
\left[\frac{4}{7} (N^2 - 1) +  2 \tilde N_F' \right]  \lmk \frac{g_*}{43/4} \rmk^{-4/3} \\
 &\simeq&
 \left[ 0.093 \lrf{N^2 - 1}{8} + 0.041 \tilde N_F' \right] \lmk \frac{g_*}{200} \rmk^{-4/3},
 \label{deltaNeff}
\eeq
where $\tilde N_F'$ denotes the flavor number of massless (or sufficiently light) hidden quarks.
Together with the SM prediction of $N_{\rm eff}^{(\rm SM)} = 3.046$, 
this is consistent with the present constraint of 
$N_{\rm eff}^{(\rm obs)} = 2.99 \pm 0.39$~\cite{Aver:2013wba, Planck:2015xua}. 
The ground-based Stage-VI CMB polarization experiment CMB-S4 
will measure the effective neutrino number with precisions of $\Delta N_{\rm eff} = 0.0156$, 
so that it can indirectly test our model~\cite{Abazajian:2013oma, Wu:2014hta}. 
In addition, the dark radiation (i.e., $SU(N)_H$ gauge boson and hidden quarks) is self-interacting in our model~\cite{Jeong:2013eza},
and so, it has different clustering properties compared to the standard free-streaming one.
The clustering properties are represented by its effective sound speed $c_{\rm eff}^2$ 
and its viscosity parameter $c_{\rm vis}^2$. 
These parameters can be measured by CMB observations~\cite{Bell:2005dr, Cirelli:2006kt, Friedland:2007vv, Smith:2011es, Archidiacono:2011gq, Diamanti:2012tg, Archidiacono:2013fha}, 
so that in principle we can distinguish between our model 
and other models that predicts free-streaming dark radiation.

\subsection{Application to the $H_u H_d$ flat direction
\label{sec:Higgs}}

In this subsection,  we consider a more economical model in which the $SU(N)_H$ gauge symmetry and 
the field $\phi$ are replaced with QCD and the $H_u H_d$ flat direction, respectively.
In this model, we will introduce
extra quark multiplets to obtain a sufficiently large dynamical scale $\Lambda_{\rm QCD}$ 
during inflation~\cite{Jeong:2013xta}. 
We consider a KSVZ-like axion model~\cite{Kim:1979if}, 
where the Higgs fields do not carry PQ charges.

Let us consider the $H_u H_d$ flat direction with a superpotential of 
\beq
 W = \frac{1}{2} \mu \phi^2 + \lambda \frac{\phi^4}{4 \Mpl},
\eeq
where we denote $\phi^2/2 \equiv H_u H_d$. 
The first term in the RHS is the usual Higgs $\mu$ term. 
This superpotential leads to the following potential of the flat direction: 
\beq
 V(\phi) = m_\phi^2 \abs{\phi}^2 
 + \frac{\lambda \mu}{\Mpl} \abs{\phi}^2 (\phi^2 + c.c. )
 - c_H H^2 \abs{\phi}^2 
 - \lmk a_H \lambda H \frac{\phi^4}{4 \Mpl} + c.c. \rmk
 + \lambda^2 \frac{\abs{\phi}^6}{\Mpl^2} 
 + V_T(\phi), 
\eeq
where $m_\phi$ ($\simeq \mu$) is the mass of the flat direction. 
Hereafter, we assume $c_H = a_H = 1$ for simplicity. 
At a finite temperature, the thermal potential $V_T$ is given as 
\beq
 V_T (\phi) \simeq 
 c_T \alpha_s^2 T^4 \log \lmk \frac{ \abs{\phi}^2 }{T^2} \rmk, 
\eeq
for $\phi \gg T/g$. 
The coefficient is given by 
$c_T = 9/8$ 
for the $H_u H_d$ flat direction. 
Since the flat direction has a tachyonic mass of order the Hubble parameter, 
it obtains the VEV of Eq.~(\ref{VEV}) during inflation and inflaton oscillation dominated era. 
Then, 
the flat direction starts to oscillate around the origin of the potential 
at the time of Eq.~(\ref{H_osc^phi}).

When the $H_u H_d$ flat direction has a large VEV during inflation, 
quark multiplets obtain effective masses much larger than the QCD scale $\Lambda_{\rm QCD}$. 
Since the renormalization group flow of the QCD coupling constant is sensitive to 
the number of light quark multiplets, 
the dynamical scale of $SU(3)_c$ depends on 
their masses, i.e., 
the VEV of the flat direction. 
A large VEV of $\phi$ can make the effective QCD scale during inflation $\Lambda_{\rm QCD}^{\rm inf}$ much larger 
than $\Lambda_{\rm QCD} \approx 400 \MEV$.
Therefore, the axion mass can be enhanced in the early Universe, and if it is heavier
than $H_{\rm inf}$ during inflation, the axion quantum fluctuations are suppressed. 
However, 
in order to make the dynamical scale as high as $10^{13} \GEV$, 
we need to introduce extra colored particles. 
We add $N'_F$ pairs of $Q'_H$ and $\bar{Q}'_H$, which 
are charged under $SU(5)$, with the following interaction,
\beq
 W_{Q'_H} = \lmk M_{Q'_H} + \frac{\phi^2}{M'} \rmk Q'_H \bar{Q}'_H.
\eeq
Note that the extra quarks do not have any PQ charges.
In this case, 
the effective QCD scale can be as high as~\cite{Jeong:2013xta}
\beq
 \Lambda_{\rm QCD}^{\rm inf} \simeq 1.3 \times 10^7 \GEV \lmk \frac{M_{\rm GUT}}{M_{Q'_H}} \rmk^{N'_F / 9}, 
\eeq
for $\phi \sim M' \sim M_{\rm GUT}$. 
As a result, the
axion mass is given by Eq.~(\ref{m_a}).%
\footnote{
In our setup, 
the up and down quarks may be lighter than $\Lambda_{\rm QCD}^{\rm inf}$. 
In this case, we should replace $\Lambda_H$ in Eq.~(\ref{m_a}) 
to $\bar{\Lambda}_H^{\rm inf}$ defined by $(\bar{\Lambda}_H^{\rm inf})^{3 N} = \Lambda_H^{3N - 2} m_u m_d$, 
where $m_u$ and $m_d$ is given by $y_u \phi$ and $y_d \phi$, respectively. 
}
If  $m_a \gtrsim H_{\rm inf}$ during inflation, 
 the isocurvature constraint can be avoided.

Next, we consider the scenario of spontaneous baryogenesis. 
As in the previous subsections, we assume $T_{\rm max} \gtrsim \Lambda_{\rm QCD}^{\rm inf}$ 
so that $SU(3)_c$ is deconfined 
and the axion becomes massless just after inflation. 
Then, when temperature decreases down to $\Lambda_{\rm QCD}^{\rm inf}$, 
the axion again obtains an effective mass due to the non-perturbative effect. 
Here, the minimum of the axion field is determined by the theta parameter 
and is given by 
\beq
 \theta_{\rm eff} = \theta_0 + \arg \lkk \det \lmk y_u y_d \rmk \rkk + 3 \arg \lkk M_{\tilde{g}} \rkk - 3 \arg \lkk \mu B \rkk, 
\eeq
where $\theta_0$ is the bare theta parameter 
and 
$M_{\tilde{g}}$ is the mass of $SU(3)_c$ gaugino. 
Since the minimum of the axion field value is related to $\theta_{\rm eff}$, 
it also depends on 
the phase of R-symmetry breaking term 
via $\arg \lkk M_{\tilde{g}} \rkk$. 
In general, the phase of R-symmetry breaking term 
changes after inflation 
because the source of R-symmetry changes after inflation. 
This implies that 
the minimum of the axion field
changes after inflation, 
so that the axion starts to oscillate at $T = \Lambda_{\rm QCD}$. 
This dynamics can be used to realize the spontaneous baryogenesis.

In contrast to the previous model,  the $SU(2)$ gauge symmetry is spontaneously broken
by the large VEV of $H_u H_d$ in the present scenario.
This implies that sphalerons are decoupled and one cannot use the
anomaly equation  to derive the lagrangian of Eq.~(\ref{anomaly eq}).
So, let us instead introduce a \Kahler potential of 
\beq
 K  &\sim& \frac{1}{f_a} (A + A^*) \lkk c^a_i \abs{\psi_i}^2 + \dots \rkk, 
\eeq
where $A$ represents the axion superfield 
and $\psi_i$ represents SM matter superfields. 
This leads to the following derivative couplings between 
axion and SM particles: 
\beq
 \mathcal{L} 
 &=& - c^a_i \frac{a(t)}{f_a} \del_\mu \lmk \bar{\psi}_i \gamma^\mu \psi_i \rmk \\
 &=& c^a_i \frac{\del_0 a}{f_a} \lmk \bar{\psi}_i \gamma^0 \psi_i \rmk + \dots. 
\eeq
This leads to an effective chemical potential for the lepton current as 
\beq
 \mu_{\rm eff} = \sum_i c^a_i g_i L_i \frac{\del_0 a}{f_a} 
 \equiv c^a \frac{\del_0 a}{f_a}, 
\eeq
where $L_i$ are lepton charges of fields $i$, 
and $g_i$ are the numbers of spin states, but with an extra factor of 2 for bosons. 
Thus the axion oscillation can induce a nonzero chemical potential for 
the lepton current. 
When we introduce a heavy right-handed neutrinos 
and realize the seasaw mechanism, 
it gives us lepton violating interactions. 
Note that 
lepton violating processes are efficient 
via the electron and higgs interactions 
though the Higgs VEV is much larger than the temperature of the Universe. 
Thus the lepton asymmetry is approximately given by Eq.~(\ref{n_L}). 
The subsequent calculation and discussion are the same with those explained in the psevious section 
and the final result is given by Eq.~(\ref{result}).

Some time after inflation,  the $H_uH_d$ flat direction starts to oscillate around the origin of the potential. 
Then, the effective QCD scale becomes equal to $\Lambda_{\rm QCD}$, and the axion becomes
massless again.
Finally, the axion starts to oscillate around the QCD phase transition and contributes to CDM.
The DM abundance can be explained when the PQ breaking scale satisfies Eq.~(\ref{f_a for DM}).

Finally, we comment on another source of baryon asymmetry in this model~\cite{Chiba:2003vp, Takahashi:2003db}. 
When the flat direction starts to oscillate at $H = H_{\rm osc}$, 
its phase direction also starts to rotate in the complex plane. 
This implies that the masses of MSSM particles 
obtain a time-dependent phase through the Yukawa interactions. 
Since the time-dependent phase of mass terms 
can be interpreted as a chemical potential, 
we obtain the chemical potential of $B+L$ current from the dynamics of the $H_u H_d$ flat direction: 
\beq
 \mu_{B + L} = 3 \omega_\phi, 
\eeq
where we define $\omega_\phi$ by $\phi = \abs{\phi} e^{i \omega_\phi t}$. 
Let us emphasize that 
the origin of this chemical potential is completely different from that considered above. 
Therefore, 
the baryon and lepton asymmetry may also be generated by the spontaneous baryogenesis 
via this chemical potential. 
However, the flat direction starts to oscillate due to the thermal log potential, 
so that the kick in the phase direction is so small that 
the rotation frequency $\omega_\phi$ is much smaller than the Hubble parameter $H_{\rm osc}$. 
This implies that the resuling chemical potential is much smaller than 
that obtained by Eq.~(\ref{mu}). 
Therefore we can neglect this contribution and justify our result of Eq.~(\ref{result}).

\section{Conclusions
\label{sec:conclusion}}

The QCD axion is one of the plausible candidates for CDM, which is, however, severely constrained
by the isocurvature perturbations. In this paper we have proposed an extension of the QCD axion
model to avoid the isocurvature constraint by suppressing  quantum fluctuations of the axion.
Specifically we have considered a scenario where the PQ symmetry is badly broken by a
larger amount in the past than in the present, due to non-perturbative effects of hidden or 
$SU(3)_c$ gauge interactions. Most importantly, the dynamical scale 
 can be temporarily enhanced during inflation, if the renormalization group flow of the gauge coupling is
 significantly modified by a flat direction with a large VEV.
If the dynamical scale is enhanced so as to make the axion mass heavier than or comparable to 
 the Hubble parameter during inflation,
the axion isocurvature perturbations are suppressed. 

The  dynamics of the axion and flat direction could be slightly involved after inflation. We 
have focused on the case in which the maximal temperature of the Universe is higher than 
the dynamical scale so that the axion becomes massless just after inflation. 
Some time after inflation, the axion becomes massive   again 
and starts to oscillate around the potential minimum when
 the temperature becomes comparable to the dynamical scale.
Interestingly, if the axion has a coupling to $SU(2)_L$ gauge fields, 
the axion oscillation induces a nonzero effective chemical potential of the $B+L$ symmetry,
which would generate the baryon/lepton asymmetry in the presence of  baryon/lepton number violating operators.
We have shown that a correct amount of the baryon asymmetry is generated by the QCD axion via spontaneous
baryogenesis, by taking account of the $\Delta L = 2$ process mediated by the heavy right-handed
Majorana neutrinos.
Soon after the baryon asymmetry is generated, 
the  flat direction starts to oscillate around the origin of the potential.
In the first model, the hidden gauge interactions then become weakly coupled and it remains so until present.
Depending on the flavor number of the hidden quarks,  the hidden gauge interactions may become asymptotic 
non-free. Thus, the axion becomes massless again.
 Finally, at the QCD phase transition, 
the axion acquires a tiny mass through non-perturbative effect of QCD instantons
and it is stabilized at the CP conserving minimum.
The  observed DM abundance can be explained by the QCD axion produced by the misalignment mechanism.
To realize the above scenario and account for  the observed baryon asymmetry and the DM abundance, 
the Hubble parameter of inflation must be of order $10^{11} \GEV$. 
We have also shown that the scenario can be similarly realized 
when the flat direction is identified with the $H_u H_d$ flat direction  once we introduce additional colored particles at an intermediate scale. 

One of the predictions of our scenario based on the hidden gauge interactions is that there must be a self-interacting
dark radiation with $\Delta N_{\rm eff} ={\cal O}(0.01-0.1)$ given by Eq.~(\ref{deltaNeff}). The ground-based Stage-VI CMB polarization 
experiment will be able to detect the dark radiation with this amount and distinguish it from free-streaming dark radiation.

In the Appendix we also consider an interaction between axion and Ricci scalar, 
which results in a heavy axion mass during inflation. 
We find that the axion abundance as well as isocurvature perturbations 
are suppressed, 
so that the axion decay constant can be as large as the GUT scale.

\vspace{1cm}

%
\section*{Acknowledgments}
This work is supported by Grant-in-Aid for Scientific research 
on Innovative Areas (No. 23104008 (F.T.)), 
Scientific Research (A) (No. 26247042 (F.T.)), 
Scientific Research (B) (No. 26287039 (F.T.)), 
JSPS Grant-in-Aid for Young Scientists (B) (No. 24740135 (F.T.)), 
JSPS Research Fellowships for Young Scientists (No. 25.8715 (M.Y.)), 
World Premier International Research Center Initiative (WPI Initiative), MEXT, Japan (F.T. and M.Y.),
and the Program for the Leading Graduate Schools, MEXT, Japan (M.Y.).

%

\vspace{1cm}

\appendix
\section{Model with a coupling to the Ricci scalar 
\label{sec:Ricci scalar}}

In this Appendix, we provide another model to suppress the axion isocurvature perturbations. 
We consider a model 
where 
the axion is coupled with the Ricci scalar as 
\beq
 \mathcal{L} = c_R^2 R \Mpl^2 \cos \lmk \frac{a}{f_a} - \theta_R \rmk, 
 \label{aR}
\eeq
where $c_R$ and $\theta_R$ are constants. 
Since $R = -6 ( (\dot{a}/a)^2 + \ddot{a}/a) \simeq - 12 H^2$ during inflation, 
the axion acquires a mass much larger than the Hubble parameter for
$c_R \gtrsim f_a / \Mpl$, and therefore the quantum fluctuations of the axion is
suppressed. 

Here we simply assume that there are no other PQ breaking terms; in particular, any
other Planck-suppressed operators are assumed to be absent or sufficiently suppressed. 
In general, the absence of such PQ breaking terms is an issue  that has been discussed
extensively in the literature~\cite{Georgi:1981pu,Lazarides:1985bj,Dine:1986bg,Kamionkowski:1992mf, 
Holman:1992us,Barr:1992qq,Dobrescu:1996jp}. Our purpose is not to
explain the long-standing problem concerning the high quality of the PQ symmetry, but to 
study cosmological impacts on such PQ breaking terms that are enhanced in the
early Universe.

The peculiarity of the above interaction (\ref{aR}) is that it suppresses not only the axion quantum fluctuations,
but also the axion DM abundance.
To see this, let us estimate the axion abundance. 
At the time around the QCD phase transition, 
the energy density of the Universe is dominated by radiation, i.e., $a \propto t^{1/2}$, 
so that 
the Ricci scalar is one-loop suppressed as~\cite{Kajantie:2002wa, Davoudiasl:2004gf}
\beq
 R &=& - 3 (1 - 3 \omega) H^2 \\
 1 - 3 \omega &\simeq& 
 \frac{162 \alpha_s^2}{19 \pi^2} + \mathcal{O}(g^5), 
\eeq
for the $SU(3)_c$ gauge theory with three flavors. 
Thus, it induces the so-called Hubble-induced term for the axion: 
\beq
 \mathcal{L} &=&
 - \frac{1}{2} c_a^2 H^2 (a - \theta_R f_a)^2 + \cdots,\\
 c_a &\simeq& 1.6 c_R \alpha_s \frac{\Mpl}{f_a},
\eeq
where we have expanded the axion potential around the potential minimum.
Since $\alpha_s \sim 1$ at the time around the QCD phase transition, 
$c_a$ is much larger than unity as long as $c_R \gg f_a / \Mpl$.
Then, the axion abundance is expected to be suppressed in a similar way as the adiabatic suppression 
mechanism for the moduli abundance proposed in Ref.~\cite{Linde:1996cx}. 

In order to estimate axion abundance, 
let us  solve the following approximated equation of motion for the axion: 
\beq
 \ddot{a} + 3 H \dot{a} \simeq - m_a^2 (T) a - c_a^2 H^2 (t) \lmk a - \theta_R f_a \rmk, 
 \label{axion EOM}
\eeq
where $H = 1/ 2t$, 
$m_a (T)$ is given by Eq.~(\ref{m_a at T ne 0}), and we have omitted higher order terms.\footnote{
This is for ease of comparison with Ref.~\cite{Linde:1996cx}.
The axion abundance can be similarly suppressed even if one solve the equation of motion without
any approximation. 
}
Let us explain the behavior of the axion qualitatively. 
First, in the regime of $H \gg m_a(T)$, 
the Hubble-induced term dominates the axion potential, 
so that the axion stays at $a \approx \theta_R f_a$. 
Then, at the time around $H \sim m_a(T) / c_a$, 
the potential minimum starts to move as $a_{\rm min}(t) \equiv \theta_R f_a c_a^2 H^2 / (m_a^2 + c_a^2 H^2)$. 
Here, the typical time scale of the shift of the potential minimum $a_{\rm min}(t)$ is of order the inverse of the Hubble parameter,
while that of the axion oscillations is of order $(c_a H)^{-1}$ ($ \ll H^{-1}$). 
This means that the minimum of the axion potential changes adiabatically,
and the axion number density in the comoving volume is the adiabatic invariant, and
practically no axion oscillations are induced through this dynamics. 
Finally, in the regime of $H \ll m_a(T) /c_a$, 
the Hubble-induced mass becomes negligible 
and the axion oscillates around the present vacuum, and its energy density decreases as $a^{-3}$.

We can estimate the abundance of the axion oscillations induced by weak
violation of the adiabaticity~\cite{future work}. 
The resulting axion abundance is roughly given by 
\beq
 \rho_a (t) \sim m_a^2 \lmk \theta_R f_a \rmk^2 \lmk \frac{t}{t_{\rm osc}} \rmk^{-3/2} e^{- 2 \pi c_a /(4+ n)}, 
\eeq
where $t_{\rm osc} \simeq c_a / 2 m_a(t_{\rm osc})$ 
for $c_a \gg 1$. 
Therefore, the axion-Ricci scalar coupling 
suppresses the axion abundance efficiently. 
The result shows that, for a certain value of $c_a$, 
the observed DM abudance can be explained 
even if the axion decay constant is as large as the GUT scale.

Finally, let us comment on the difference of the present model from the original 
one in Ref.~\cite{Linde:1996cx}, where the moduli mass in the low energy is
constant with time. In fact, it was pointed out in Refs.~\cite{Nakayama:2011wqa, Nakayama:2011zy}
that the adiabaticity of the moduli dynamics is necessarily broken by the
inflaton dynamics, leading to a non-negligible production of the moduli oscillations.
This is because, while the inflaton is lighter than the modulus field which acquires a mass 
of ${\cal O}(10-100) H_{\rm inf}$ during inflation, the inflaton eventually becomes heavier than the 
modulus field some time after inflation. When the two masses are comparable, the adiabaticity
is necessarily broken, which leads to production of some amount of moduli oscillations.
In particular, the moduli abundance is only power suppressed and not exponentially
suppressed. In contrast, the axion abundance is exponentially suppressed in our scenario,
because the axion mass $m_a(T)$ is negligible until the QCD phase transition.
In the context of moduli problem, our result implies that the adiabatic suppression mechanism
works successfully  and the moduli abundance is exponentially suppressed, if the moduli
potential in the low energy vanishes during and some time after inflation and arises at a sufficiently late time.



\end{document}